\documentstyle[aps]{revtex}

\begin{document}
\draft
\title{Pair Correlations in Scale-Free Networks}
\author{Huang Zhuang-Xiong, Wang Xin-Ran and Zhu Han\thanks{%
Author to whom correspondence should be addressed; Email:
hanzhunju@hotmail.com}}
\address{Department of Physics, Nanjing University, Nanjing 210093, China}
\date{\today }
\maketitle

\begin{abstract}
Correlation between nodes is found to be a common and important property in
many complex networks. Here we investigate degree correlations of the
Barabasi-Albert (BA) Scale-Free model with both analytical results and
simulations, and find two neighboring regions, a disassortative one for low
degrees and a neutral one for high degrees. The average degree of the
neighbors of a randomly picked node is expected to diverge in the limit of
infinite network size. As an generalization of the concept of correlation,
we also study the correlations of other scalar properties, including age and
clustering coefficient. Finally we propose a correlation measurement in
bipartite networks.

\end{abstract}

\pacs{PACC: 0250, 0520}

Complex networks, described by a set of nodes and links between them, play
an important role in the understanding of many natural systems. They take
various forms in different fields, including Internet\cite
{internet,internet-corr}, social networks\cite{social}, ecological networks%
\cite{ecol}, etc. (See Ref. \cite{rev1,rev2} for review.) Yet due to
technical difficulties of information acquisition and analysis, the
prosperity of research in this area did not come until only several years
ago, mainly after the finding that complex networks in reality usually do
not conform to the long assumed theory of random network proposed by Erdos
and Renyi\cite{er}.

Among the models aimed at explaining the observed nontrivial structures, the
Barabasi and Albert (BA) Scale-Free model\cite{ba} is well known for
describing systems where geographical distance is not so important. It
introduces two mechanisms that are believed to be common in reality: network
growth (adding one node at each time step and connecting it to $m$ existing
nodes) and preferential attachment (well-connected nodes tend to receive
more new links, with the probability proportional to the number of the links
they already have). It has explained the power-law degree distribution
widely observed in reality (degree of a node is defined as the number of its
nearest neighbors). Based on it some new mechanisms have been further
introduced to describe other observations such as aging effect\cite
{class-smw}, high clustering, and hierarchical structure\cite{hi}, etc.

An important property of networks, and the topic of the present article, is
the pair correlation (or mixing pattern)\cite
{internet,internet-corr,corr-ran,ass-mix,mix-pat,between}. ''Whether
gregarious people are more likely to contact with gregarious people?''\ or
''Whether an old web site is more likely to connect to old ones?''\ These
questions often have fundamental importance in reality. The topology and
many properties such as resilience and percolation of a network are all
susceptible to the answers. Very recently, a particular kind of it, the pair
degree correlation, has received much attention. Newman has put forward a
direct measurement\cite{ass-mix,mix-pat}, making it a quantified property,
and has applied this measurement to various real networks. The data
available reveal an interesting feature: in social networks, high degree
nodes often tend to be connected to other high degree nodes (assortative
mixing), while in technological and biological networks high degree nodes
often tend to connect to low degree nodes (disassortative mixing). At the
same time there are some systems reported to show no significant biases
(neutral mixing). This assortative (disassortative, or neutral) pattern has
a profound effect on the structure of evolving networks. For example, it may
tend to break a network into separate communities and affect its
error-and-attack tolerance, etc.

As a leading model in this field of complex networks, the BA model\cite{ba}
has been developed and studied extensively in the past few years (see Ref. 
\cite{rev1,cal,exact} and references therein). Yet as to correlation,
theoretical investigation of the BA model is still inadequate, which we
believe is fundamental to understand the correlation properties of real
networks, since many of them are shown to be scale free. In this sense, the
present work on the pair-correlation properties of the BA model may provide
a basis for theoretical explanations of the observations in reality.
Especially, this model is likely to be a good starting point for finding the
origin of the emergence of mixing patterns.

In this article, we focus on the correlation properties within the framework
of the BA networks. We provide analytical calculation of the average degree
of a node's nearest neighbors, indicating that the BA networks are neutral
in the limit of large connectivity, otherwise they are disassortative. The
comparison of analytical result and simulation shows that the absolute value
of the average connectivity of neighboring nodes also depends on the size of
the network and diverges as this size tends to infinity. We then extend the
investigation to the correlation of age and clustering coefficient, i.e. the
ratio of edges among the nearest neighbors of a selected node and the
maximal number of edges among them. The measurements of these two properties
have not been carried out as far as our knowledge goes. In the final part,
we propose a way to measure mixing patterns in bipartite networks.

In the following we call a node with $k$ degrees a $D-k$ node, a node with
age $a$ an $A-a$ node, a node born at time $t$ a $T-t$ node, and a node with
clustering coefficient $c$ a $C-c$ node.

{\it Degree Correlation}: The best way to completely describe degree
correlation is to obtain the matrix, $P\left( k,k_{nn}\right) $, i.e. the
probability that a nearest neighbor of a $D-k$ node is $D-k_{nn}$. However
due to technical problems this measurement is often unfeasible in practice%
\cite{internet,internet-corr}. Instead Pastor-Satorras {\it et al}. have
suggested measurement of $\left\langle k_{nn}\right\rangle _k$\cite
{internet-corr}, i.e., the average degree of a nearest neighbor of the $D-k$
nodes. Here our work about the BA model is along this line.

In Ref. \cite{cal}, Krapivsky and Render has obtained in the BA model a
useful characterization of correlation, $N_{kl}\left( t\right) $, the number
of $D-k$ nodes that attach to a $D-l$ ancestor. Asymptotically, $%
N_{kl}\left( t\right) \rightarrow tn_{kl}$, where 
\begin{equation}
n_{kl}=\frac{4\left( l-1\right) }{k\left( k+1\right) \left( k+l\right)
\left( k+l+1\right) \left( k+l+2\right) }+\frac{12\left( l-1\right) }{%
k\left( k+l-1\right) \left( k+l\right) \left( k+l+1\right) \left(
k+l+2\right) }.
\end{equation}
With this result Krapivsky and Render concluded that the BA networks have an
assortative mixing pattern\cite{cal}, i.e., nodes with similar degrees are
more likely to be connected. Their calculation has considered the direction
of links between nodes, which makes the measurement a rather complex task in
practice. Here, as in almost all measurements carried out so far, we
consider undirected links. Similarly we study $N_{kl}^{\prime
}=N_{kl}+N_{lk} $, the number of connected pairs of nodes with $k$ degrees
and $l$ degrees respectively. Asymptotically 
\[
N_{kl}^{\prime }/t\rightarrow n_{lk}^{\prime }=n_{kl}+n_{lk}. 
\]
The probability that a nearest neighbor of a $D-k$ node is $D-k_{nn}$ is 
\begin{equation}
\left( N_{k,k_{nn}}+N_{k_{nn},k}\right) /\sum_{k_{nn}}\left(
N_{k,k_{nn}}+N_{k_{nn},k}\right)
\end{equation}
and the average degree of a nearest neighbor of a $D-k$ node is 
\begin{equation}
\left\langle k_{nn}\right\rangle _k=\frac{\sum_{k_{nn}}k_{nn}\left(
N_{k,k_{nn}}+N_{k_{nn},k}\right) }{\sum_{k_{nn}}\left(
N_{k,k_{nn}}+N_{k_{nn},k}\right) }\rightarrow \frac{\sum_{k_{nn}}k_{nn}%
\left( n_{k,k_{nn}}+n_{k_{nn},k}\right) }{\sum_{k_{nn}}\left(
n_{k,k_{nn}}+n_{k_{nn},k}\right) }.
\end{equation}
This is a useful result characterizing the degree correlation of the BA
networks. We show it approximately in Fig. 1(a) by taking the summation of $%
l $ to $1.5\times 10^5$ and $2\times 10^6$ (with normalization satisfied)
respectively, in comparison with simulation results at $N=100,1000$ and $%
10000$. (The simulation results reported here are all obtained by taking
average of up to $5000$ independent runs). It reveals that nodes with large $%
k$ show no obvious biases in their associations, as has been reported in
previous studies. However, when $k$ is relatively small, $\left\langle
k_{nn}\right\rangle _k$ falls significantly as $k$ increases, and this is a
sign of disassortative mixing. (This region has not been noticed previously
and, actually, {\it nodes falling in this region often take up a significant
part in real networks that have been measured so far}.) In Fig. 1(b) we show
two characteristic distributions of the probability that a nearest neighbor
of a $D-k$ node is $D-k_{nn}$. Both simulation and theoretical results show
that this probability declines as approximately $k_{nn}^{-2}$ for large $%
k_{nn}$, which is a sign of neutral mixing since the probability of finding
a $D-k_{nn}$ node in the network is approximately $k_{nn}^{-3}$\cite{ba}. It
also leads to the conclusion that $\left\langle k_{nn}\right\rangle $ will
diverge for infinite network size $N$ as $\ln N$ (since the largest possible
value of $k\sim N^{1/2}$, see Eq. (\ref{kwithN}) below), and this trend is
observed in Fig. 1(a). The deviation from this power law is evident when $%
k_{nn}$ is small. On the other hand, as can be noticed, the simulation and
theoretical result agree better for larger $k$. Here the finite size effect
mainly results from the fact that, when a network has a finite size, the
nodes with degrees out of a certain range will be absent. Finally, Fig. 1(c)
shows the simulation result of the probability matrix, $P\left(
k,k_{nn}\right) $. While for each value of $k$ the probability has similar
distribution for large $k_{nn}$ ($k_{nn}^{-2}$), the difference is obvious
for relatively small values of $k_{nn}$.

{\it Age Correlation}: As an extension of the degree correlation, we define
age correlation in a similar way, i.e. consider the probability that an $%
A-a_i$ node is linked with an $A-a_j$ node. (Here age $a$ of a node is
defined as $t_F-t$, where $t_F$ is the age of the network and $t$ denotes
the time when the node is born.) If this conditional probability $P\left(
a,a_{nn}\right) $ is independent of $a$, we are in a topology without any
age correlation. We may also study the quantity $\left\langle
a_{nn}\right\rangle _a$, i.e. the average age of a nearest neighbor of the $%
A-a$ nodes. For convenience, in the following we shall calculate $%
\left\langle t_{nn}\right\rangle _t$ instead of $\left\langle
a_{nn}\right\rangle _a$.

In a BA network, at each time step, a newly born node is connected to $m$
existing nodes, while the rate at which a node receives new links is
proportional to its current degree. Using continuous approach\cite{ba}, we
have 
\begin{equation}
\frac{dk_s\left( t\right) }{dt}=\frac{mk_s\left( t\right) }{2mt}=\frac{%
k_s\left( t\right) }{2t},
\end{equation}
where $k_s\left( t\right) $ denotes the degree of a node born at time $s$.
Solving this equation we can get

\begin{equation}
k_s\left( t\right) =m\left( \frac ts\right) ^{\frac 12}.  \label{kwithN}
\end{equation}
Consider two nodes born at time $t_i$ and $t_j$ respectively. Assume $%
t_i<t_j $, and the probability that they are connected 
\begin{equation}
p\left( t_i,t_j\right) =\frac{mk_{t_i}\left( t_j\right) }{2mt_j}=\frac m{2%
\sqrt{t_it_j}},  \label{a-mattrix}
\end{equation}
where $k_{t_i}\left( t_j\right) $ is the degree of the $T-t_i$ node at time $%
t_j$. This result is also valid if $t_i>t_j$. Then 
\begin{equation}
\left\langle t_{nn}\right\rangle _{t_i}=\frac{\sum_{t_j=1}^{t_F}t_jp\left(
t_i,t_j\right) }{\sum_{t_j=1}^{t_F}p\left( t_i,t_j\right) }=\frac{%
\sum_{t_j=1}^{t_F}\sqrt{t_j/t_i}}{\sum_{t_j=1}^{t_F}\left( 1/\sqrt{t_it_j}%
\right) }.
\end{equation}
Replace the summation with integration, and we have 
\begin{equation}
\left\langle t_{nn}\right\rangle _{t_i}\approx \frac{\int_1^{t_F}\sqrt{%
t_j/t_i}dt_j}{\int_1^{t_F}\left( 1/\sqrt{t_it_j}\right) dt_j}=\frac 13\frac{%
t_F^{3/2}-1}{t_F^{1/2}-1}\approx \frac 13t_F,  \label{tnn-average}
\end{equation}
which is independent of $t_i$. To support the above analysis, we show
simulation results in Fig 2. $\left\langle t_{nn}\right\rangle _t$ as a
function of $t$ is shown in Fig. 2(a), and is found to be fluctuating around 
$t_F/3$, as predicted by Eq. (\ref{tnn-average}). In Fig. 2(a), the
nearest-neighbor age distributions of three characteristic nodes all show an
power law, $t_{nn}^{-\gamma }$, with the exponent $\gamma \approx -0.5$, in
accordance with Eq. (\ref{a-mattrix}). Fig. 2(c) shows the matrix $P\left(
t,t_{nn}\right) $, i.e. the probability that a nearest neighbor of a $T-t$
node is $T-t_{nn}$. We can see clearly from this graph that the distribution
is largely independent of $t$, which indicates a neutral pattern in this age
correlation.

{\it Clustering Correlation}: As another extension of this concept, we also
investigate pair correlation of the clustering coefficient, using the same
method as that in the preceding parts. Different from a node's degree,
clustering coefficient is a continuous parameter between $0$ and $1$. In
order to define the correlation, discretization is necessary. Here we evenly
divide the region $\left[ 0,1\right] $ into $250$ subregions and each
subregion is represented by the median value. In this way, the correlation
of clustering can be defined in the way similar as before, i.e., consider
the probability that a node with its clustering coefficient in the $c-$%
subregion (for convenience we call it a $C-c$ node) is linked with an $%
C-c_{nn}$ node. If this conditional probability $P\left( c,c_{nn}\right) $
is independent of $c$, we are in a topology without any correlation.

In our simulation, the clustering coefficients are mainly limited in a
relatively small region $\left[ 0,0.14\right] $ (we can see from Fig. 3(b)),
and we have poor statistics concerning the nodes out of this region. Our
simulation result (Fig. 3(a)) shows that the average clustering coefficient
of the nearest neighbors of a $C-c$ node, $\left\langle c_{nn}\right\rangle
_c$, generally shows a trend to ascend as $c$ increases. This is a sign of
assortative mixing in the sense of clustering, which may be because that
clustering is a collective behavior. If a node with large $c$ belongs to a
cluster, it is likely that its neighbors also belong to that cluster and
hence also have large $c$. At the same time, we find that, interestingly,
there is a notable peak in the curve of $\left\langle c_{nn}\right\rangle _c$%
. We may call it a central mixing pattern, in order to develope the ideas of
Newman\cite{ass-mix,mix-pat}. The value of this peak agrees with that of the
clustering coefficient distribution of the network. This coincidence
suggests that the observed peak may result from the fact that, in the
network, the nodes that have their clustering coefficients around this value
take up a significant part of the system. Fig. 3(b) shows the $P\left(
c,c_{nn}\right) $ distributions with three typical values of $c$. Three
curves all show a peak at approximately the above-mentioned position---a
coincidence again. Fig. 3(c) shows the non-trivial part of the matrix $%
P\left( c,c_{nn}\right) $.

Before conclusion we briefly discuss the correlation measurement in
bipartite networks\cite{rev2,bipart}, which are formed by two (or more)
classes $A$ and $B$ of nodes with links running between only different
kinds. They may describe various systems, e.g. books and readers in a
library, or directors and boards in the business world\cite{rev2}. Here a
new kind of correlation may be of major interest, the correlation between
the properties of an $A$ node and the same properties of the $B$ nodes it is
linked with. And this may answer such questions as whether a much-reading
reader borrows books that are also liked by other people.

To conclude, as is now widely accepted, correlation is an important index of
and significantly influences network structure and function, in that it
reveals the relationship between a node and its neighborhood. This article
concerns theoretically with the correlation of the BA model. Most of the
previous studies focus on the degree correlation, and here we generalize the
concept to various properties. We analytically treat the degree correlation
(disassortative plus neutral) and the age correlation (neutral) of the BA
network, and provide simulation results about the clustering coefficient
correlation (assortative plus central). The mixing patterns of these scalar
properties and their origins are presented and explained. Considering the
wide applicability and theoretical importance of the BA model, the results
reported here may serve as a useful reference for further studies,
especially for the investigation in the variants of the BA model that
incorporate, for example, nonlinear preferential attachment\cite
{nonlinear-pre}, aging effect\cite{class-smw}, etc.

Presently the observed assortative (disassortative) patterns in reality have
stimulated efforts to study their influence on network structure and
function, mainly with models that are created by adding correlation effect
to random networks\cite{corr-ran,ass-mix,mix-pat}. However, it is well known
that real networks have robust organization principles (for example, the
preferential attachment, as characterized by the BA model). What is the
relationship between correlation and these principles? What is the origin of
the correlation pattern in the process of network growth? These questions
are far from being completely answered. Recent efforts\cite{origin} have
partly shown how they can be related to some features added to the BA model,
such as fitness and some growing restraints.

We appreciate helpful discussions with A.-L. Barabasi and K. Klemm. This
work was partially supported by the National Science Fund for Distinguished
Young Scholars of China under Grant No. 60225014.

\null\vskip0.2cm

\centerline{\bf Caption of figures} \vskip1cm

Fig. $1$. Simulation results of degree correlation of a network with $m=3$.
(a) Average degree $\left\langle k_{nn}\right\rangle $ of the neighboring
nodes of the $D-k$ nodes as a function of $k$. Squares, circles and upward
triangles represent the simulation results, with system size $N=100$, $1000$
and $10000$ respectively. Downward triangles and diamonds represent the
theoretical result with $l$ up to $1.5\times 10^5$ and $2\times 10^6$
respectively. (b) Degree distributions of the nearest neighbors of $D-3$
nodes (squares) and $D-20$ nodes (circles) respectively. The dashed lines
are the corresponding theoretical results. The solid line with slope $-2$
serves as a guide to the eye. (c) The probability matrix $P\left(
k,k_{nn}\right) $.

Fig. $2$. Simulation results of age correlation of a network with size $%
N=10^4$, $t_F=500$ and $m=3$. (a) Average introduction time of the nearest
neighbors of a $T-t$ node as a function of $t$. (b) Introduction time
distributions of the neighbors of the $T-1$ (squares), $T-100$ (circles) and 
$T-200$ (upward triangles) node. The solid line has the slope $-0.5$. (c)
The matrix $P\left( t,t_{nn}\right) $.

Fig. $3$. Simulation results of clustering coefficient correlation of a
network with size $N=2500$ and $m=25$. (a) Average clustering coefficient of
the neighbors of $C-c$ nodes as a function of $c$. (b) Clustering
coefficient distributions of the nearest neighbors of the $C-0.034$
(squares), $C-0.054$ (circles) and $C-0.074$ (upward triangles) nodes. The
distribution of clustering coefficient of the network (multiplied by $1.68$
for convenience) is shown as stars. (c) The matrix $P\left( c,c_{nn}\right) $%
.

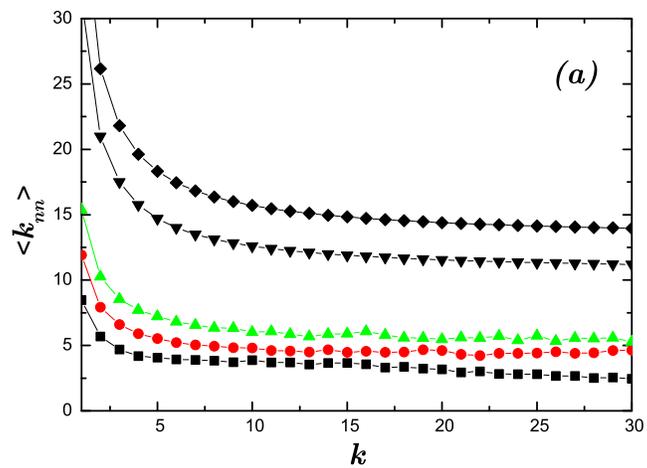
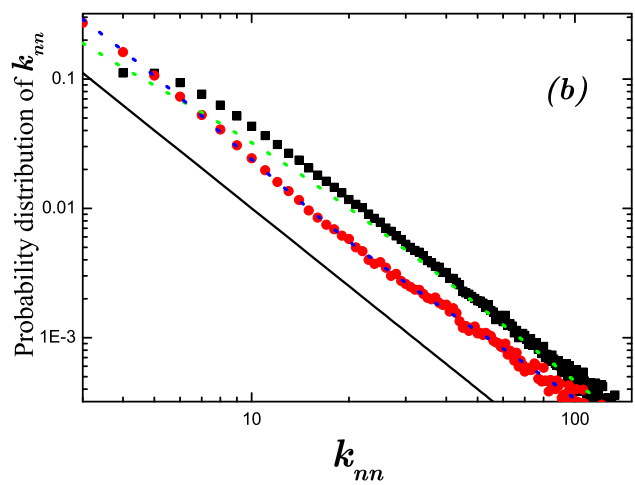
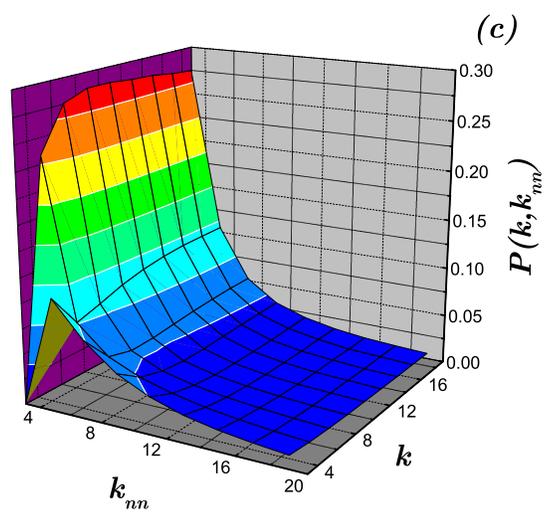
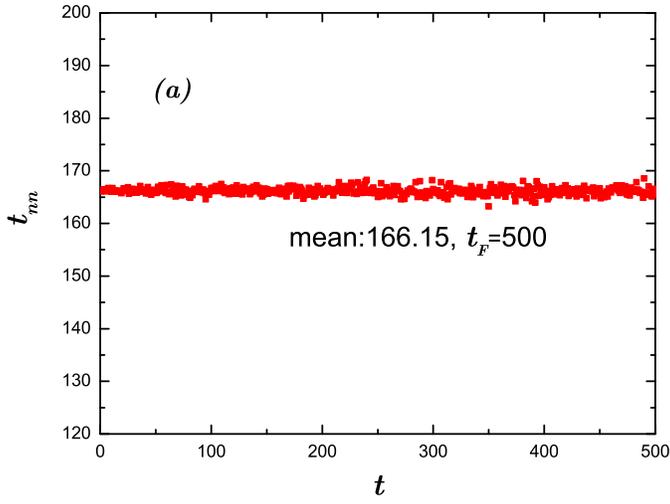
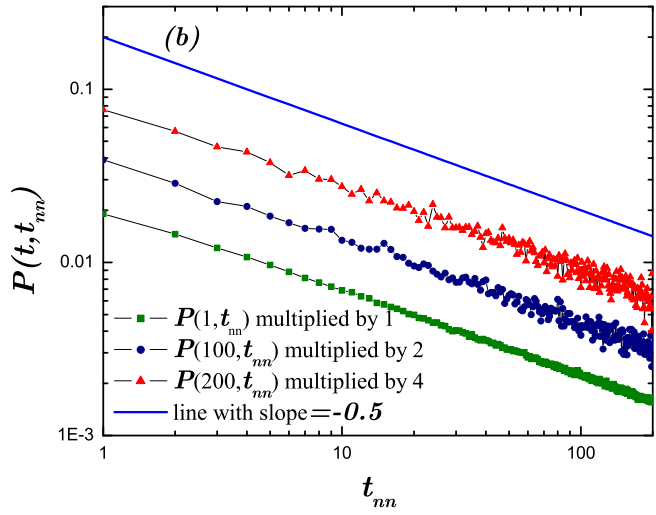
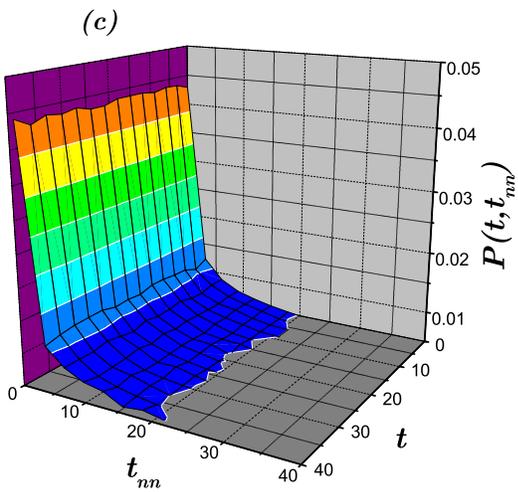
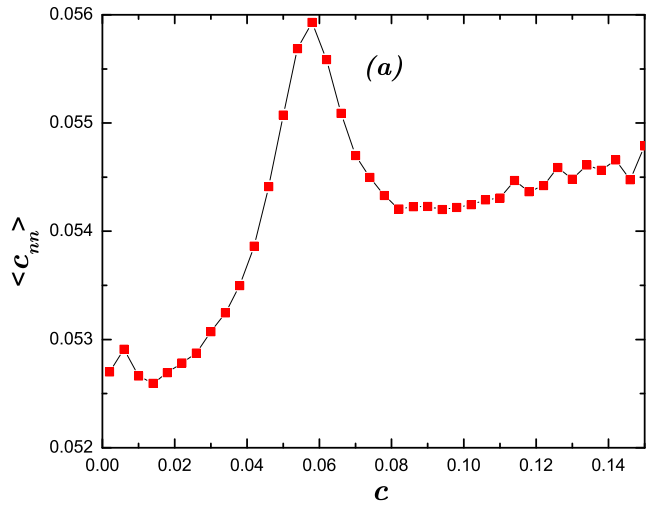
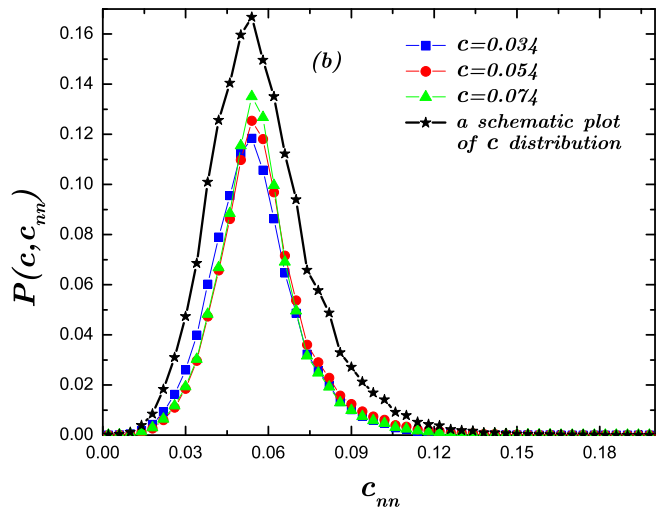
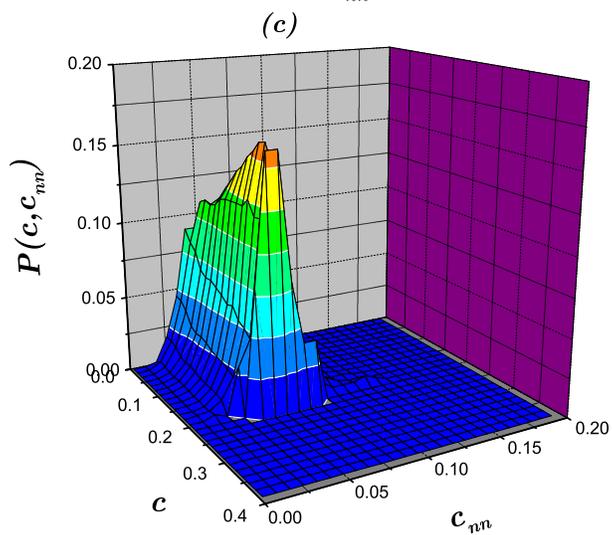
\end{document}